\documentclass[preprint,showpacs,preprintnumbers,amsmath,amssymb]{revtex4}
\usepackage{graphicx}

\begin{document}

\thispagestyle{empty}

\title{The impact of magnetic properties on atom-wall
interaction}

\author{G.~Bimonte,${}^1$
G.~L.~Klimchitskaya,${}^2$ and
V.~M.~Mostepanenko${}^3$}

\affiliation{${}^1$Dipartimento
di Scienze Fisiche, Universit\`{a}
di Napoli Federico II, Complesso
Universitario MSA, Via Cintia
I---80126 Napoli, Italy \\
INFN, Sezione di
Napoli, Napoli, Italy \\
${}^2${North-West Technical University,
Millionnaya St. 5, St.Petersburg,
191065, Russia}\\
${}^3${Noncommercial Partnership ``Scientific Instruments'',
Tverskaya St. 11, Moscow,
103905, Russia}
}

\begin{abstract}
The Lifshitz-type formulas for the free energy and Casimir-Polder
force acting between an atom possessing a permanent magnetic
moment and a wall made of different materials are derived.
Simple model allowing analytic results is considered where the
atomic magnetic susceptibility is frequency-independent and wall
is made of ideal metal. Numerical computations of the Casimir-Polder 
force are performed for H atom interacting with walls made of
ideal metal, nonmagnetic (Au) and ferromagnetic (Fe)
metals and of ferromagnetic dielectric. It is shown
that for the first three wall materials the inclusion of the
magnetic moment of an atom decreases and for the fourth material
increases the magnitude of the Casimir-Polder force.
\end{abstract}
\maketitle

\section{Introduction}

During the last few years atom-wall interaction attracts considerable
attention in connection with experiments on quantum reflection
\cite{1,2,3} and Bose-Einstein condensation \cite{4,5,6}. 
This stimulated investigation of the van der Waals and Casimir-Polder
atom-wall interaction potentials including their dependence on atomic
and wall properties \cite{7,8,9,10,11}. Most of previous work was
devoted to a pure electrically polarizable atom near a metallic or
dielectric wall described by the frequency-dependent dielectric
permittivity $\varepsilon$ and equal to unity magnetic permeability
$\mu$. However, the role of magnetic properties of an atom and wall
material has also been discussed. Magnetic properties received much
interest due to highly conjectural possibility of repulsive atom-wall
interaction based on the finding \cite{12} that a pure electrically
polarizable atom repels a pure magnetically polarizable one. 
Keeping in mind that both atoms and walls used in cavity quantum
electrodynamics may possess magnetic proiperties, the impact of these
properties on atom-wall interaction deserves detailed consideration.

Recent Ref.~\cite{13} developed the theory of atom-wall interaction
for the case of both a polarizable and a (para)magnetizable atom
near a magnetodielectric macrobody. This theory was applied to the
case of an atom near a semispace (thick magnetodielectric wall
described by the frequency-dependent $\varepsilon$ and $\mu$).
It was shown that the resulting potential of atom-wall interaction
is very similar to the known respective potential of a polarizable
atom interacting with a dielectric wall.
It is pertinent to note that Ref.~\cite{13} deals with paramagnetic
atoms which are magnetizable but have no intrinsic magnetic moment.
This is what is referred to as the Van Vleck paramagnetism \cite{14}.
It is caused be the deformation of the electron structure of an atom 
by the external field which creates the induced magnetic moment.
Usually such deformation leads to the diamagnetic effect. However,
in some specific cases the paramagnetic affect arises \cite{14}.
Thus, the Van Vleck paramagnetism is of polarization origin and
the respective magnetic susceptibility is temperature-independent.

In this paper we consider the impact of magnetic properties on atom-wall
interaction for paramagnetic atoms possessing the intrinsic
(permanent) magnetic moment. Such atoms (for instance H or Rb)
participate in different physical processes involving atom-surface
interaction (see, e.g., Refs.~\cite{15,16}). As wall material, we present
computations for a nonmagnetic metal and ferromagnetic metal and
dielectric. The case of atoms possessing a permanent magnetic moment is 
interesting in two aspects. First, the magnitude of a permanent
magnetic moment is much larger than the magnitude of an induced one.
Second, the resulting magnetic susceptibility is of orientation
origin and it is temperature-dependent. In Sec.~II we derive the
Lifshitz-type formula for an atom with permanent magnetic moment
interacting with a magnetodielectric wall starting from the known
formula for two semispaces described by $\varepsilon(\omega)$ and
$\mu(\omega)$. Section III is devoted to the case of nonmagnetic
wall. We begin with simple model of an ideal metal wall and
frequency-independent magnetic susceptibility of an atom and
demonstrate that for Rb atom the effect of magnetic moment is
negligibly small because the electric polarizability remains much
larger than the magnetic susceptibility at all temperatures from 1 to 300\,K.
For H atom the inclusion of magnetic moment leads to minor {\it decrease} of
the magnitude of atom-wall force at $T=1\,$K. Then, similar computations
are performed for H atom near an Au wall. In Sec.~IV the case of H atom
near walls made of ferromagnetic materials is considered. It is shown
that for ferromagnetic metal (Fe) the inclusion of atomic magnetic moment
leads to qualitatively the same result as for nonmagnetic metals.
For ferromagnetic dielectrics the inclusion of atomic magnetic moment
{\it increases} the magnitude of atom-wall force at $T=1\,$K where
effect of magnetic properties is most pronounced. Section V is devoted to
our conclusions and discussion.

\section{Lifshitz-type formula for atom-wall interaction with
account of magnetic properties}

We start from the Lifshitz formula for the free energy per unit area
in configuration of two parallel magnetodielectric semispaces
separated by a distance $a$, at temperature $T$ in thermal equilibrium
\cite{21}
\begin{eqnarray}
&&{\cal{F}}(a,T)=\frac{k_BT}{2\pi}
\sum\limits_{l=0}^{\infty}{\vphantom{\sum}}^{\prime}
\int_{0}^{\infty}k_{\bot}dk_{\bot}\left\{\ln\left[1-
r_{\rm TM}^{(1)}({\rm i}\xi_l,k_{\bot})r_{\rm TM}^{(2)}({\rm i}\xi_l,k_{\bot})
e^{-2aq_l}\right]\right.
\nonumber \\
&&\phantom{aaaaa}+\left.\ln\left[1-
r_{\rm TE}^{(1)}({\rm i}\xi_l,k_{\bot})r_{\rm TE}^{(2)}({\rm i}\xi_l,k_{\bot})
{\rm e}^{-2aq_l}\right]\right\}.
\label{eq1}
\end{eqnarray}
\noindent
Here the reflection coefficients $r_{\rm TM,TE}^{(n)}$
for two magnetodielectric semispaces ($n=1,\,2$) are given by
\cite{17,18}
\begin{eqnarray}
&&r_{\rm TM}^{(n)}({\rm i}\xi_l,k_{\bot})=
\frac{\varepsilon_l^{(n)}q({\rm i}\xi_l,k_{\bot})-
k^{(n)}({\rm i}\xi_l,k_{\bot})}{\varepsilon_l^{(n)}q({\rm i}\xi_l,k_{\bot})
+k^{(n)}({\rm i}\xi_l,k_{\bot})},
\nonumber \\
&&r_{\rm TE}^{(n)}({\rm i}\xi_l,k_{\bot})=\frac{\mu_l^{(n)}
q({\rm i}\xi_l,k_{\bot})-
k^{(n)}({\rm i}\xi_l,k_{\bot})}{\mu_l^{(n)}q({\rm i}\xi_l,k_{\bot})
+k^{(n)}({\rm i}\xi_l,k_{\bot})},
\label{eq2}
\end{eqnarray}
\noindent
and the dielectric permittivities and magnetic permeabilities, 
$\varepsilon_l^{(n)}\equiv\varepsilon^{(n)}({\rm i}\xi_l)$,
$\mu_l^{(n)}\equiv\mu^{(n)}({\rm i}\xi_l)$,
are calculated at the imaginary Matsubara frequencies,
$\xi_l=2\pi k_BTl/\hbar$ where $l=0,\,1,\,2,\ldots\,$, $k_B$ is the Boltzmann
constant. The other notations in Eqs.~(\ref{eq1}), (\ref{eq2}) are
as follows:
\begin{eqnarray}
&&
q_l\equiv q({\rm i}\xi_l,k_{\bot})
=\sqrt{k_{\bot}^2+\frac{\xi_l^2}{c^2}},
\label{eq3} \\
&&
k_l^{(n)}\equiv k^{(n)}({\rm i}\xi_l,k_{\bot})
=\sqrt{k_{\bot}^2+\varepsilon_l^{(n)}\mu_l^{(n)}\frac{\xi_l^2}{c^2}},
\nonumber
\end{eqnarray}
\noindent
$k_{\bot}$ is the wave vector projection onto the boundary planes 
restricting both
semispaces. A prime near the summation sign means that the term for
$l=0$ has to be multiplied by 1/2.

At both zero and nonzero temperature, Eq.~(\ref{eq1}) with
frequency-independent $\varepsilon^{(n)}$ and $\mu^{(n)}$ was used
to determine the values of $\varepsilon^{(n)}$ and $\mu^{(n)}$
leading to a positive energy (free energy) and respective
repulsive Casimir force between two magnetodielectric
semispaces \cite{19}. It was shown \cite{20}, however,
that in the range of frequencies which give the major contribution
to the Casimir force $\mu$ is nearly equal to unity far away from
the values needed for the realization of Casimir repulsion.

In order to obtain the free energy of a magnetic atom near a
magnetodielectric semispace we use the same method as was
suggested for the electrically polarizable atom near a dielectric
semispace \cite{21}. For this purpose we remain the semispace with 
$n=1$ unchanged but consider a rarefied magnetodielectric semispace with
$n=2$ as a paramagnetic gas. Expanding the dielectric permittivity
and magnetic permeability of the latter in powers of the number of atoms 
per unit volume $N$ and preserving only the first-order contributions one
obtains \cite{21,22}
\begin{eqnarray}
&&
\varepsilon_l^{(2)}=1+4\pi\Gamma({\rm i}\xi_l)=
1+4\pi N\alpha({\rm i}\xi_l)+O\left(N^2\right),
\nonumber \\
&&
\mu_l^{(2)}=1+4\pi\chi({\rm i}\xi_l)=
1+4\pi N\beta({\rm i}\xi_l)+O\left(N^2\right).
\label{eq4}
\end{eqnarray}
\noindent
Here, $\Gamma({\rm i}\xi_l)$, $\chi({\rm i}\xi_l)$ are dynamic 
electric and magnetic susceptibilities of the rarified material
of the semispace with $n=2$,
$\alpha({\rm i}\xi_l)$ and $\beta({\rm i}\xi_l)$ are the respective
quantities, as applied to one atom. It should be remembered that
the quantity $\alpha({\rm i}\xi_l)$ is usually temperature-independent,
whereas $\beta({\rm i}\xi_l)$ for paramagnetic materials with the
orientation polarization is the reciprocal to the temperature
(see Sec.~III).

Substituting Eq.~(\ref{eq4}) in Eq.~(\ref{eq2}), using Eq.~(\ref{eq3}) and
expanding up to the first power in $N$, one arrives at
\begin{eqnarray}
&&
r_{\rm TM}^{(2)}({\rm i}\xi_l,k_{\bot})=
\pi N\left[2\alpha({\rm i}\xi_l)-\frac{\beta({\rm i}\xi_l)+
\alpha({\rm i}\xi_l)}{q_l^2}\,\frac{\xi_l^2}{c^2}\right]
+O\left(N^2\right),
\nonumber \\
&&
r_{\rm TE}^{(2)}({\rm i}\xi_l,k_{\bot})=
\pi N\left[2\beta({\rm i}\xi_l)-\frac{\beta({\rm i}\xi_l)
+\alpha({\rm i}\xi_l)}{q_l^2}\,\frac{\xi_l^2}{c^2}\right]
+O\left(N^2\right).
\label{eq5}
\end{eqnarray} 
\noindent
Rewriting Eq.~(\ref{eq1}) in terms of the result (\ref{eq5})
we obtain
\begin{eqnarray}
&&{\cal{F}}(a,T)=-\frac{k_BTN}{2}
\sum\limits_{l=0}^{\infty}{\vphantom{\sum}}^{\prime}
\int_{0}^{\infty}k_{\bot}dk_{\bot}{\rm e}^{-2aq_l}
\label{eq6} \\
&&~~~~~~\times
\left\{\left[2\alpha({\rm i}\xi_l)-\frac{\beta({\rm i}\xi_l)+
\alpha({\rm i}\xi_l)}{q_l^2}\,\frac{\xi_l^2}{c^2}\right]
\,r_{\rm TM}^{(1)}({\rm i}\xi_l,k_{\bot})\right.
\nonumber \\
&&~~~~~~~~
+\left.
\left[2\beta({\rm i}\xi_l)-\frac{\beta({\rm i}\xi_l)
+\alpha({\rm i}\xi_l)}{q_l^2}\,\frac{\xi_l^2}{c^2}\right]
\,r_{\rm TE}^{(1)}({\rm i}\xi_l,k_{\bot})\right\}
+O\left(N^2\right).
\nonumber
\end{eqnarray} 
\noindent
Alternatively, the additivity of the first order term in the expansion
of the free energy in powers of $N$ results in
\begin{equation}
{\cal{F}}(a,T)=N
\int_{a}^{\infty}{\cal{F}}^{A}(z,T)dz
+O\left(N^2\right),
\label{eq7}
\end{equation}
\noindent
where ${\cal{F}}^{A}(z,T)$ is the free energy of 
a magnetic atom spaced $z$ apart of a magnetodielectric  wall.

Now we equate the right-hand sides of Eqs.~(\ref{eq6}) and (\ref{eq7})
and calculate the derivative with respect to $a$. Then in the limit
$N\to 0$ we obtain
\begin{eqnarray}
&&{\cal{F}}^{A}(a,T)=-k_BT
\sum\limits_{l=0}^{\infty}{\vphantom{\sum}}^{\prime}
\int_{0}^{\infty}k_{\bot}dk_{\bot}q_l{\rm e}^{-2aq_l}
\label{eq8} \\
&&\phantom{aaaa}\times
\left\{
\vphantom{\frac{\xi_l^2}{q_l^2c^2}}
2[\alpha({\rm i}\xi_l)r_{\rm TM}^{(1)}({\rm i}\xi_l,k_{\bot})
+\beta({\rm i}\xi_l)r_{\rm TE}^{(1)}({\rm i}\xi_l,k_{\bot})]\right.
\nonumber \\
&&\phantom{aaaa}\left.
-\frac{\xi_l^2}{q_l^2c^2}[\alpha({\rm i}\xi_l)+
\beta({\rm i}\xi_l)]\,
[r_{\rm TM}^{(1)}({\rm i}\xi_l,k_{\bot})+
r_{\rm TE}^{(1)}({\rm i}\xi_l,k_{\bot})]\right\}.
\nonumber
\end{eqnarray}
\noindent
At zero temperature similar formula for the energy of a magnetizable
atom was obtained in Ref.~\cite{13} using the Green function method.
For a nonmagnetic atom, $\beta({\rm i}\xi_l)=0$, near a dielectric
wall, $\mu({\rm i}\xi_l)=1$, Eq.~(\ref{eq8}) coincides with the
results of Refs.~\cite{7,21} (one should take into account different
convention for the phase multiple in the definition of the TE reflection
coefficient used in \cite{7}).

{}From Eq.~(\ref{eq8}) it is straightforward matter to derive the
expression for the force acting on a magnetic atom near a
magnetodielectric wall
\begin{eqnarray}
&&
F^{A}(a,T)=-\frac{\partial{\cal{F}}^{A}(a,T)}{\partial a}
=-2k_BT
\sum\limits_{l=0}^{\infty}{\vphantom{\sum}}^{\prime}
\int_{0}^{\infty}k_{\bot}dk_{\bot}q_l^2{\rm e}^{-2aq_l}
\label{eq9} \\
&&\phantom{aaaa}\times
\left\{
\vphantom{\frac{\xi_l^2}{q_l^2c^2}}
2[\alpha({\rm i}\xi_l)r_{\rm TM}^{(1)}({\rm i}\xi_l,k_{\bot})
+\beta({\rm i}\xi_l)r_{\rm TE}^{(1)}({\rm i}\xi_l,k_{\bot})]\right.
\nonumber \\
&&\phantom{aaaa}\left.
-\frac{\xi_l^2}{q_l^2c^2}[\alpha({\rm i}\xi_l)+
\beta({\rm i}\xi_l)]\,
[r_{\rm TM}^{(1)}({\rm i}\xi_l,k_{\bot})+
r_{\rm TE}^{(1)}({\rm i}\xi_l,k_{\bot})]\right\}.
\nonumber
\end{eqnarray}

It is interesting to note that both the free energy (\ref{eq8}) and
the force (\ref{eq9}) of atom-wall interaction are represented as the 
sums of two contributions
\begin{eqnarray}
&&
{\cal F}^A(a,T)={\cal F}_{\alpha}^A(a,T)+{\cal F}_{\beta}^A(a,T),
\nonumber \\
&&
F^A(a,T)=F_{\alpha}^A(a,T)+F_{\beta}^A(a,T),
\label{eq10}
\end{eqnarray}
\noindent
depending on the dynamic atomic polarizability $\alpha$ and
magnetic susceptibility $\beta$, respectively. However, magnetic
properties of wall material influence on both contributions to
the free energy and force through the magnetic permeability $\mu$
entering the reflection coefficients $r_{\rm TM,TE}^{(1)}$
defined in Eq.~(\ref{eq2}). In the next section, Eq.~(\ref{eq9})
is used in numerical computations to determine the impact of
magnetic properties on atom-wall interaction.

\section{Atoms with permanent magnetic moment near a 
\protect{\\} nonmagnetic wall}

To perform computations of the force acting between an atom and a wall
using Eq.~(\ref{eq9}), one needs sufficiently precise expressions
for the atomic dynamic polarizability $\alpha$ and magnetic
susceptibility $\beta$. For a rarefied gas of paramagnetic atoms the
magnetic susceptibility along the imaginary frequency axis is given
by \cite{23,24}
\begin{equation}
\chi({\rm i}\xi_l)=N\beta({\rm i}\xi_l)=N
\,\frac{g^2\mu_B^2J(J+1)}{3k_BT}\,\frac{1}{1+\tau\xi_l},
\label{eq11}
\end{equation}
\noindent
where $g$ is the Lande factor, $\mu_B=e\hbar/(2m_ec)$ is the Bohr
magneton, $m_e$ is the electron mass, $J$ is the total momentum and
$\tau$ is the relaxation time. Below we consider the ground state
atoms of H and ${}^{87}$Rb which have approximately equal magnetic
moments \cite{25}. For these atoms $g=1$ and $J=1/2$ (the magnetic moment
of H atoms was determined in Ref.~\cite{26}; the relativistic and
radiative corrections to it are discussed in Ref.~\cite{27}).
For different atoms at $T=300\,$K, $\tau$ varies in the range from
$10^{-10}$ to $10^{-4}\,$s and increases with the decrease of temperature.

Now we consider the dynamic atomic polarizability. As is seen from
Eq.~(\ref{eq11}), at high frequencies the magnetic properties cannot
have a pronounced effect on the force between an atom and a cavity wall.
Because of this, below we consider the impact of magnetic properties
on the Casimir-Polder force in the separation range from 1 to 10$\,\mu$m
where the relevant frequencies are relatively low. In this range of
separations sufficiently precise results for the Casimir-Polder force
are obtained by using the single-oscillator model \cite{7}
\begin{equation}
\alpha({\rm i}\xi_l)=\frac{\alpha(0)}{1+\frac{\xi_l^2}{\omega_a^2}},
\label{eq12}
\end{equation}
\noindent
where $\alpha(0)$ is the static atomic polarizability and $\omega_a$ is the
eigenfrequency. For H atom 
$\alpha(0)=6.67\times 10^{-25}\,\mbox{cm}^3$ and the characteristic
energy is $\hbar\omega_a=11.65\,$eV \cite{28}. For ${}^{87}$Rb atom it holds
$\alpha(0)=4.73\times 10^{-23}\,\mbox{cm}^3$ \cite{29} and the characteristic
energy is $\hbar\omega_a=1.68\,$eV \cite{30}.

In this section we consider the case of nonmagnetic metal walls.
Note that for nonmagnetic dielectric walls the magnetic moment of an
atom leaves the Casimir-Polder force unaffected. This is because the
magnetic susceptibility (\ref{eq11}) is dominant at zero frequency.
It is well known \cite{31}, however, that for nonmagnetic dielectrics
$r_{\rm TE}^{(1)}(0,k_{\bot})=0$. Thus, from Eqs.~(\ref{eq8}) and (\ref{eq9})
it follows that for nonmagnetic dielectrics there is no impact of the
atomic magnetic moment on atom-wall interaction. The case of walls made of
magnetic materials is considered in the next section.

It is more convenient to perform computations by using the dimensionless
variables
\begin{equation}
\zeta_l=\frac{2a\xi_l}{c}=\frac{\xi_l}{\omega_c},\quad
y=2aq_l,
\label{eq13}
\end{equation}
\noindent
where $\omega_c\equiv c/(2a)$ is the characteristic frequency of the
Casimir-Polder interaction. In terms of these variables Eq.~(\ref{eq9})
takes the form
\begin{eqnarray}
&&
F^{A}(a,T)=-\frac{k_BT}{8a^4}
\sum\limits_{l=0}^{\infty}{\vphantom{\sum}}^{\prime}
\int_{\zeta_l}^{\infty}ydy{\rm e}^{-y}
\left\{
2y^2\left[\alpha({\rm i}\omega_c\zeta_l)
r_{\rm TM}^{(1)}({\rm i}\omega_c\zeta_l,y)\right.\right.
\nonumber \\
&&~~~~\left.+\beta({\rm i}\omega_c\zeta_l)
r_{\rm TE}^{(1)}({\rm i}\omega_c\zeta_l,y)\right]-
\zeta_l^2\left[\alpha({\rm i}\omega_c\zeta_l)+
\beta({\rm i}\omega_c\zeta_l)\right]
\nonumber \\
&&~~~~\left.
\times
\left[r_{\rm TM}^{(1)}({\rm i}\omega_c\zeta_l,y)+
r_{\rm TE}^{(1)}({\rm i}\omega_c\zeta_l,y)\right]\right\},
\label{eq14}
\end{eqnarray}
where the reflection coefficients (\ref{eq12}) are
\begin{eqnarray}
&&
r_{\rm TM}^{(1)}({\rm i}\omega_c\zeta_l,y)=
\frac{\varepsilon_l^{(1)}y-\sqrt{y^2+\zeta_l^2(\varepsilon_l^{(1)}
\mu_l^{(1)}-1)}}{\varepsilon_l^{(1)}y+\sqrt{y^2+\zeta_l^2(\varepsilon_l^{(1)}
\mu_l^{(1)}-1)}},
\nonumber \\
&&
r_{\rm TE}^{(1)}({\rm i}\omega_c\zeta_l,y)=
\frac{\mu_l^{(1)}y-\sqrt{y^2+\zeta_l^2(\varepsilon_l^{(1)}
\mu_l^{(1)}-1)}}{\mu_l^{(1)}y+\sqrt{y^2+\zeta_l^2(\varepsilon_l^{(1)}
\mu_l^{(1)}-1)}}
\label{eq15}
\end{eqnarray}
\noindent
with $\varepsilon_l^{(1)}=\varepsilon^{(1)}({\rm i}\omega_c\zeta_l)$,
$\mu_l^{(1)}=\mu^{(1)}({\rm i}\omega_c\zeta_l)$ 

As a simple model, we first consider atom with frequency-independent
electric polarizability $\alpha(0)$ and magnetic susceptibility
$\beta(0)$ near an ideal metal wall. Then from Eq.~(\ref{eq15})
one obtains $r_{\rm TM}^{(1)}=1$, $r_{\rm TE}^{(1)}=-1$  and
Eq.~(\ref{eq14}) results in
\begin{equation}
F^A(a,T)=-\frac{k_BT}{4a^4}[\alpha(0)-\beta(0)]
\sum\limits_{l=0}^{\infty}{\vphantom{\sum}}^{\prime}
\int_{\zeta_l}^{\infty}y^3{\rm e}^{-y}dy.
\label{eq16}
\end{equation}
\noindent
The calculation of the integral in Eq.~(\ref{eq16}) leads to
\begin{equation}
F^A(a,T)=-\frac{k_BT}{4a^4}[\alpha(0)-\beta(0)]\,
\left[3+\sum\limits_{l=1}^{\infty}(6+6\zeta_l+3\zeta_l^2+\zeta_l^3)
{\rm e}^{-\zeta_l}\right].
\label{eq17}
\end{equation}
\noindent
By performing all summations in Eq.~(\ref{eq17}) one obtains
\begin{eqnarray}
&&
F^A(a,T)=-\frac{k_BT}{4a^4}[\alpha(0)-\beta(0)]\,
\left[
\vphantom{\frac{({\rm e}^{2\tau})}{({\rm e}^{\tau}-1)^4}}
3+\frac{6}{{\rm e}^{\tau}-1}+\frac{6\tau}{({\rm e}^{\tau}-1)^2}\right.
\nonumber \\
&&~~~~~
\left.
+\frac{3\tau^2{\rm e}^{\tau}(1+{\rm e}^{\tau})}{({\rm e}^{\tau}-1)^3}
+\frac{\tau^3{\rm e}^{\tau}(1+4{\rm e}^{\tau}+
{\rm e}^{2\tau})}{({\rm e}^{\tau}-1)^4}\right],
\label{eq18}
\end{eqnarray}
\noindent
where the parameter $\tau$ has the meaning of the normalized
temperature $\tau=2\pi T/T_{\rm eff}$, and the effective temperature
is defined as $k_BT_{\rm eff}=\hbar c/(2a)$.

Equations (\ref{eq18}) shows that the impact of atomic magnetic moment
on atom-wall interaction is determined by the relationship between
the static electric polarizability $\alpha(0)$ and static atomic
susceptibility $\beta(0)$. From Eq.~(\ref{eq11}) one arrives at the
following values for $\beta(0)$ of H and ${}^{87}$Rb atoms at
$T=300\,$K and $T=1\,$K, respectively:
$\beta(0;T=300\,K)=5.2\times 10^{-28}\,\mbox{cm}^3$,
$\beta(0;T=1\,K)=1.56\times 10^{-25}\,\mbox{cm}^3$.
From the comparison with the values of $\alpha(0)$ for H and ${}^{87}$Rb
atoms presented below Eq.~(\ref{eq12}) it follows that the impact
of atomic magnetic moment on the interaction of Rb atoms with
a cavity wall is negligibly small. We emphasize that this conclusion
is obtained in the model where $\beta$ is frequency-independent
and equal to its static value. The more so in situations when the
decrease of $\beta$ with increasing frequency is taken into 
consideration. As to H atoms, the impact of their magnetic moment
is also negligibly small at $T=300\,$K but is comparable
with the role of electric polarizability at $T=1\,$K keeping in mind
that $\beta(0)=0.23\alpha(0)$. Here we do not consider very low
temperatures $T\ll 1\,$K, where $\beta(0)$ might become even larger
than $\alpha(0)$ as is suggested by Eq.~(\ref{eq11}). The reason is that
at very low temperature even weak magnetic interaction between
separate atoms in the rarefied paramagnetic gaseous medium influences
on its magnetic properties and makes Eq.~(\ref{eq11}) inapplicable
\cite{32}. Thus, the case of very low temperature deserves further
investigation.

Now we present the results of numerical computations in more realistic 
situations. We begin with the case of H atom characterized by the
frequency-dependent $\alpha({\rm i}\xi_l)$ and  $\beta({\rm i}\xi_l)$
interacting with an ideal metal wall. In this case Eq.~(\ref{eq14})
results in
\begin{equation}
F^A(a,T)=-\frac{k_BT}{4a^4}
\sum\limits_{l=0}^{\infty}{\vphantom{\sum}}^{\prime}
\int_{\zeta_l}^{\infty}y^3{\rm e}^{-y}dy
[\alpha({\rm i}\omega_c\zeta_l)-\beta({\rm i}\omega_c\zeta_l)].
\label{eq19}
\end{equation}
\noindent
The computations using Eqs.~(\ref{eq11}), (\ref{eq12}) and (\ref{eq19})
were performed at $T=1\,$K at separations from 1 to $10\,\mu$m.
In Eq.~(\ref{eq11}) the value $\tau=10^{-8}\,$s was used. It was checked 
that further increase of $\tau$ does not influence the force values.
The computationsl results for the magnitude of the Casimir-Polder
force multipled by the fifth power of separation are presented in
Fig.~1a. The solid line reproduces conventional results for $F_{\alpha}^A$
obtained by discarding the magnetic moment of H atom [i.e., by
assuming $\beta({\rm i}\omega_c\zeta_l)=0$]. The dotted line represents
the computational results for $F^A$ with account of both dynamic
electric polarizability and magnetic susceptibility of H atom.
The relative deviation between the results of two computations,
$(|F^A|-|F_{\alpha}^A|)/|F_{\alpha}^A|$, is equal to --0.018\% at
the shortest separation $a=1\,\mu$m and achieves --0.18\% at
$a=10\,\mu$m.

For  H atom with 
frequency-dependent $\alpha({\rm i}\xi_l)$ and  $\beta({\rm i}\xi_l)$
near an Au wall computations were performed using
Eqs.~(\ref{eq11}), (\ref{eq12}), (\ref{eq14})  and Eq.~(\ref{eq15})
with $\mu_l^{(1)}=1$. For the dielectric permittivity of Au the
plasma model
\begin{equation}
\varepsilon({\rm i}\xi_l)=1+\frac{\omega_p^2}{\xi_l^2},
\label{eq20}
\end{equation}
\noindent
where $\omega_p=9.0\,$eV is the plasma frequency, has been used.
As was shown in Ref.~\cite{33}, at separations $a>400\,$nm the
description of the dielectric properties of Au by means of the plasma
model is very accurate.
The computational results for $a^5|F^A|$
at $T=1\,$K in the separation range from 1 to $10\,\mu$m are shown in
Fig.~1b (the solid line is for $\beta=0$ and the dotted line is for
both $\alpha$ and $\beta$ not equal to zero). The relative deviation
between the two lines varies from --0.015\% to --0.15\% when
separation increases from 1 to $10\,\mu$m.
As is seen from the comparison of Fig.~1a and Fig.~1b, the correction
due to the nonzero skin depth only quantitatively influences the
computational results leaving the relative difference between the solid and 
dotted lines nearly unchanged.

\section{Atoms with permanent magnetic moment near walls
made of ferromagnetic materials}

Now we consider the cavity wall made of ferromagnetic materials which are
characterized by rather large magnetic permeabilities $\mu$ \cite{22,23,24}.
It is common knowledge that in this case $\mu$ is a function of the
strength of the magnetic field {\boldmath$H$} and achieves maximum values
at rather large $|\mbox{\boldmath$H$}|$. 
As to the frequency-dependence of $\mu$,
it is of the same form as in Eq.~(\ref{eq11}), but with much larger values 
of $\tau$ than for a paramagnetic gas. Because of this, the influence
of magnetic properties of ferromagnetic materials on atom-wall
interaction occurs through the contribution of the zero-frequency term
of the Lifshitz formula. Notice that the Lifshitz formulas (\ref{eq1}),
(\ref{eq8}) and (\ref{eq9}) were derived using the standard expression
for the magnetic induction
$\mbox{\boldmath$B$}=\mu(\omega)\mbox{\boldmath$H$}$, where the magnetic
susceptibility depends only on frequency. Thus, in the applications
of the Lifshitz formula to ferromagnetic materials it is justified to
use $\mu(\omega)$ from the initial point
($\mbox{\boldmath$B$}=\mbox{\boldmath$H$}=0$) of the normal magnetization
curve where these assumptions are satisfied.

Let us consider the interaction of H atom with Fe wall. The computations
of the Casimir-Polder force were performed using Eqs.~(\ref{eq14}) and 
(\ref{eq15}). For wall material the values
$\mu(0)=1000$ and $\hbar\omega_p=11.1\,$eV \cite{34} were used.
The computational results for $a^5|F^A(a,T)|$ at $T=1\,$K are
presented in Fig.~2a (with the same notations for the solid and dotted
lines as in Fig.~1). In the case of Fe wall the relative contribution of
atomic magnetic moment at the shortest separation $a=1\,\mu$m is equal
to only $-8\times 10^{-5}$\%. At $a=10\,\mu$m it increases till --0.13\%.
Thus, for metal walls the inclusion of ferromagnetic properties does not
increase the role of magnetic moment of an atom in atom-wall
interaction.

Now we turn to the consideration of walls made of ferromagnetic
dielectrics. These are composite materials having physical properties typical
for dielectrics, but demonstrating ferromagnetic behavior under the
influence of external magnetic field. One example is a substance on the
basis of a magnetically soft iron powder and a polymer compound. 
Recently it was suggested to design such materials on the basis of air-stable
iron-cobalt nanoparticles \cite{35}. Ferromagnetic dielectrics are
widely used in magnetooptic waveguides (see, e.g., Ref.~\cite{36} and 
references therein). In the case of ferromagnetic dielectrics,
the value of the TE reflection coefficient at zero frequency is
approximately equal to unity, whereas the TM reflection coefficient
remains to be less than unity.

As an example we perform computations of the Casimir-Polder force
acting between H atom and a wall made of polyethylene with 
$\varepsilon(0)=3$ with a fraction of iron powder. For the magnetic
permeability of such compaund material $\mu(0)=100$ was used.
Computations were performed using Eq.~(\ref{eq14}) in the separation
region from 1 to $10\,\mu$m (note that at such large separations the
frequency dependence of the dielectric permittivity does not contribute
essentially to the obtained results).
The computational results for $a^5|F^A(a,T)|$ at $T=1\,$K are
presented in Fig.~2b by the solid line (with atomic magnetic properties
discarded) and by the dotted line (with atomic magnetic properties
included). In qualitative difference with the case of nonmagnetic and
ferromagnetic metal wall materials (see Fig.~1a,b and Fig.~2a), for
a wall made of ferromagnetic dielectric the inclusion of atomic magnetic
moment increases the magnitude of the Casimir-Polder force.
At the shortest separation $a=1\,\mu$m the realtive deviation between
the computational results with included and discarded atomic magnetic
moment is equal to 0.04\%. At separation distance of $a=10\,\mu$m
this deviation achieves 0.4\%. Thus, for wall made of ferromagnetic
dielectric the correction to the Casimir-Polder interaction due to the 
atomic magnetic moment is larger than for other wall materials
considered above.

\section{Conclusions and discussion}

In the foregoing we have derived the Lifshitz-type formulas for the
Casimir-Polder free energy and force acting between the atom with
a permanent magnetic moment and a wall made of different materials.
These formulas express the free energy and force in terms of electric
polarizability and magnetic susceptibility of an atom, and dielectric
permittivity and magnetic permeability of a wall. Using a simple 
model of the atom with frequency-independent  electric
polarizability and magnetic susceptibility near an ideal metal wall
the analytical expression for the Casimir-Polder force was obtained.
Specifically, for Rb atoms the influence of magnetic properties
on the force was shown to be negligibly small, as compared with H
atoms. We have also performed numerical computations of the
Casimir-Polder force acting between H atoms with frequency-dependent
electric polarizability and magnetic susceptibility and walls made
of ideal metal, Au, Fe and ferromagnetic dielectric.
In the first three cases the inclusion of an atomic magnetic moment
was shown to lead to the decrease of the force magnitude and
in the fourth case to the increase of it. Although the impact of the
permanent magnetic moment of an atom on atom-wall interaction was
found to be equal to only a fraction of percent, it is larger than
the effect of the induced (para)magnetic moment previously considered
in the literature \cite{13}.

Note that our analysis is not applicable to atoms under the influence
of an external magnetic field. This is because the inclusion of
the magnetic field changes the mathematical expression for the Casimir
force between the plates used as a starting point in Sec.\ II \cite{37,38}.
In fact the most interesting configuration considered above is
the H atom near a ferromagnetic dielectric wall. The point is that for
metal walls there are supplementary magnetic interactions caused by
the magnetic noise from Johnson currents \cite{39,40}. 
In the case of an atom interacting with a dielectric wall there
is no action of such effects which makes this configuration
preferable for further investigations.

\section*{Acknowledgments}

 G. Bimonte was partially supported by the
PRIN
SINTESI.
G.\ L.\ Klimchitskaya and V.\ M.\ Mostepanenko were supported by
the
University of Naples Federico II, under the program
"Programma
Internazionale di Modilit$\grave{{\rm a}}$ Docenti e Studenti",
and by the PRIN
SINTESI.
They were also partially supported by
Deutsche
Forschungsgemeinschaft grant
436\,RUS\,113/789/0--4.



\begin{figure*}[h]
\vspace*{.7cm}
\centerline{
\includegraphics{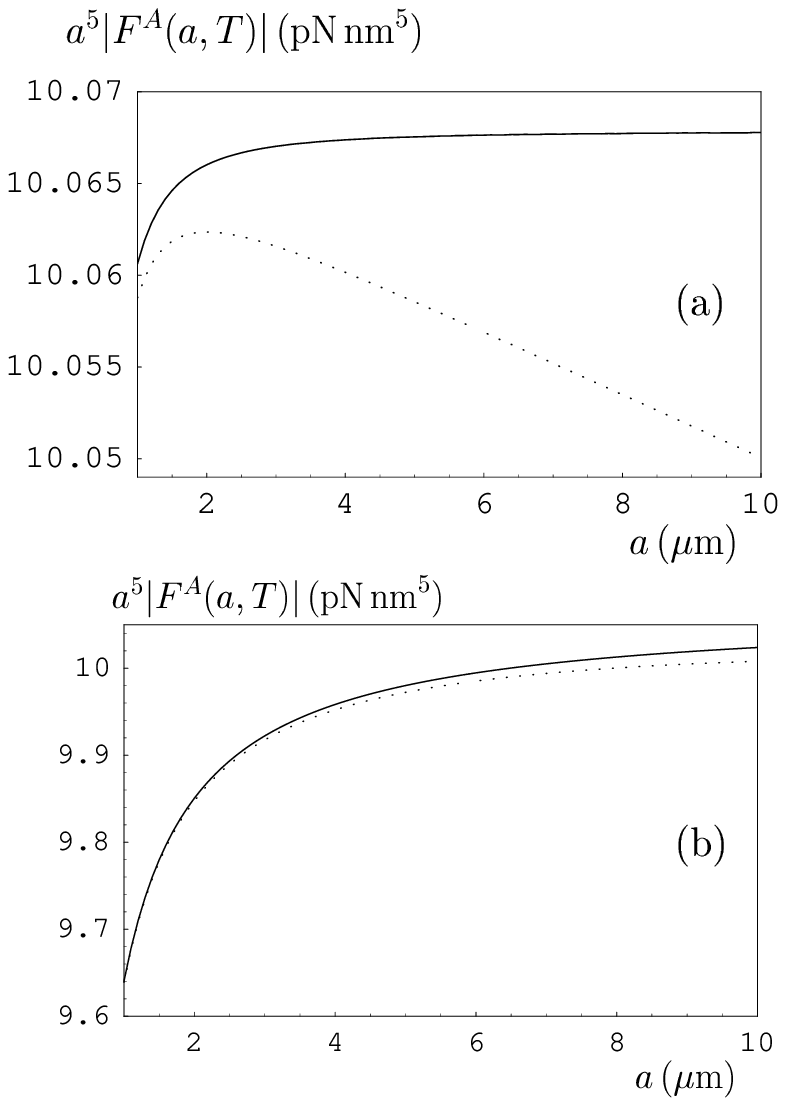}
}
\vspace*{-16.3cm}
\caption{The magnitude of the Casimir-Polder force acting
between H atom and (a) ideal metal and (b) nonmagnetic
metal (Au) wall multiplied by
the fifth power of separation with discarded (the solid line)
and included (the dotted line) atomic magnetic moment.}
\end{figure*}
\begin{figure*}[h]
\vspace*{2.cm}
\centerline{
\includegraphics{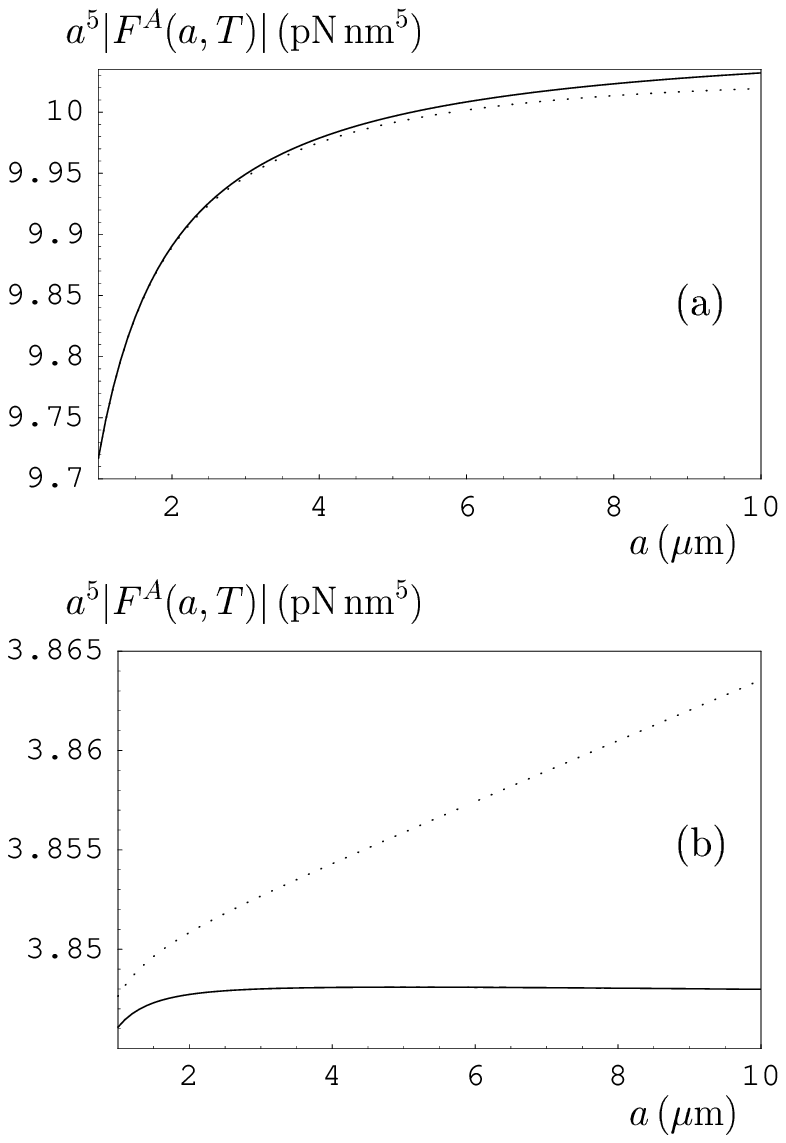}
}
\vspace*{-16.3cm}
\caption{The magnitude of the Casimir-Polder force acting
between H atom and (a) ferromagnetic metal (Fe) and
(b) ferromagnetic dielectric wall multiplied by
the fifth power of separation with discarded (the solid line)
and included (the dotted line) atomic magnetic moment. }
\end{figure*}

\end{document}